\documentclass
{mn2e}
\usepackage{epsfig}
\usepackage{amsmath, amssymb,bm}

\def \be{\begin{equation}}
\def \ee{\end{equation}}
\def \rsun{\rm R_{\odot}}
\def \msun{\rm M_{\odot}}

\begin{document}
\title[Quasi--Periodic Eruptions from Galaxy Nuclei]{Quasi--Periodic Eruptions from Galaxy Nuclei}
\author[Andrew King] 
{\parbox{5in}{Andrew King$^{1, 2, 3}$ 
}
\vspace{0.1in} \\ $^1$ School of Physics \& Astronomy, University
of Leicester, Leicester LE1 7RH UK\\ 
$^2$ Astronomical Institute Anton Pannekoek, University of Amsterdam, Science Park 904, NL-1098 XH Amsterdam, The Netherlands \\
$^{3}$ Leiden Observatory, Leiden University, Niels Bohrweg 2, NL-2333 CA Leiden, The Netherlands}

\maketitle
\begin{abstract}
I consider quasi--periodic eruptions (QPEs) from galaxy nuclei. All the known cases fit naturally into a picture of accretion from white dwarfs (WDs) in highly eccentric orbits about the central black holes which decay through gravitational wave emission. I argue that ESO 243-39 HLX--1 is a QPE source at an earlier stages of this evolution, with correspondingly a longer period, more extreme eccentricity, and significantly more massive WD donor. I show explicitly that mass transfer in QPE systems is always highly stable, despite recent claims to the contrary in the literature. This stability may explain the alternating long--short eruptions seen in some QPE sources. As the WD orbit decays, the eruptions occupy larger fractions of the orbit and become brighter, making searches for quasi--periodicities in bright low--mass galaxy nuclei potentially fruitful. 
\end{abstract}



\section{Introduction} \label{sec:intro}

The nuclei of several galaxies are observed to produce quasi--periodic eruptions (QPEs: Miniutti et al., 2019; Giustini et al. 2020; Song et al. 2020;
Payne et al. 2020;
Arcodia et al. 2021; Chakraborty et al. 2021). 
Their characteristic feature is outbursts by factors $\sim 100$ in X--rays. 
At least five systems of this type are recognised, and I shall argue here that there is a sixth previously unrecognised QPE system. It is very likely that the list of QPE sources will continue to grow as a result of searches of archived X--ray data for periodicities.  

In most currently recognised QPE sources the outbursts last of order $\sim 1$~hr and recur with rough periodicities of a few hours, but sources with longer outburst and recurrence times (up to $\sim 100$~d and $\sim  1$~yr respectively) are beginning to appear (see Table 1). In many cases the X--rays have  ultrasoft blackbody spectra with peak temperatures $T \sim 10^6$~K and luminosities $ L \sim 5\times 10^{42}\,{\rm erg\, s}^{-1}$. These imply blackbody radii $R_{\rm bb} \sim 10^{10}$~cm, a little larger than the gravitational radius 
$R_g = GM_1/c^2 = 6 \times 10^{10}M_5$~cm of a massive black hole (MBH) of mass $M_1 \sim 10^5M_5\msun$, such as may be present in low--mass galaxy nuclei.

The large amplitudes and (in the first five recognised sources) the short eruption timescales, strongly suggest repeating mass transfer events. The most natural way to reproduce the short duty cycles is to assume that gas overflows 
the tidal lobe of a star in a strongly eccentric orbit about the central black hole which is losing orbital angular momentum to gravitational radiation. The overflowing gas must form an accretion disc to produce the observed X--ray emission, and the periodic injections of more gas, or the presence of the companion near pericentre, may cause this to accrete rapidly. 

I consider this type of model here. In the rest of this paper I use the word `binary' to denote a system consisting of the galaxy's central black hole plus an orbiting donor star. I do not exclude the possibility of multiple donors orbiting the same central black hole -- pericentre passages are short
compared with the orbital periods, so there would in general be no interaction between the donors, and we can treat each `binary' independently.

In King (2020) I presented a binary model for the first QPE system (GSN 069, Miniutti et al., 2019), where it is apparent that the only 
reasonable candidate for the orbiting donor is a low--mass white dwarf. Observational selection makes this natural: main sequence donors fill their tidal lobes in much wider orbits, making the gravitational wave (GW) losses and mass transfer rates smaller (cf eqn \ref{MS} below) and so producing lower luminosities. Things are still worse for giant donors, and neutron--star donors are evidently much rarer. 

A consequence of this model is that the accreting matter may show evidence of CNO processing, and this is indeed found in GSN~069 (Sheng et al., 2021). Chen et al. (2021) applied this binary model to the five then recognised QPE sources (also correcting an error in King (2020) which did not have serious consequences), and found acceptable fits to low--mass white dwarfs in all cases (the first five entries of Table 1). The required eccentricities for these systems are in the range $0.901 < e < 0.972$, and the white dwarf masses $M_2 = m_2\msun$ are in the range $0.15 < m_2 < 0.46$. 

This paper has four main goals. First, the treatments by King (2020) and Chen et al. (2021) both assume low masses for the white dwarf donors. There is no obvious reason to exclude white dwarfs of any mass up to the Chandrasekhar limit, and I include them analytically here. 

Second, I extend this picture to show that  the well--observed system ESO 243-39 HLX--1
is probably a QPE source. This must have a long and extremely eccentric orbit, and the white dwarf donor may have a larger mass than in the short--period sources. Its relatively irregular light curve may result because the orbital period is long enough that the accretion disc is depleted from time to time.

Third, as first noted by Minutti et al. (2019) for GSN~069, the eruptions in some QPE systems often appear to show an alternating character of long and short recurrence times, which correlate in a complex way with their amplitudes. This might be taken as evidence for an alternative picture in which a star on an eccentric orbit plunges through a pre--existing accretion disc around the central black hole, even though not all of the QPE galaxy nuclei are known to be active and therefore must possess a disc of this kind (see Section \ref{sec:discussion} for further discussion of this). Here I
instead suggest that this alternation may be a generic property of WD binary models for bright QPEs because mass transfer is stable, despite statements to the contrary appearing in the literature. 

Finally, I consider the evolution of QPE binaries. It appears that they begin mass transfer with long orbital periods and extreme eccentricities. This may result via direct capture from single--star scattering, or possibly from the Hills mechanism (Cupari et al., 2022).
 Although mass transfer complicates the mathematics, the subsequent orbital evolution of QPE binaries under gravitational radiation is qualitatively similar to the basic picture found by Peters (1964) for detached binaries. Both the period and eccentricity decrease, until at short periods there is a tendency for
the systems to become less recognisable as QPE sources as the mass transfer is spread more evenly around the orbit.

\section{Mass Transfer}
\label{sec:transfer}

The paper by King (2020) found a low--mass white dwarf donor for GSN 069. Chen et al. (2021) applied this model 
systematically to 
four systems discovered subsequently, and I largely follow their treatment here. But instead of assuming that the white dwarf has low mass, so that its radius $R_2 = r_2\rsun$ varies with is mass $M_2$ as $M_2^{-1/3}$, I allow for the full range by adopting the analytic fit of Nauenberg (1972):
\be{}
r_2 = 0.01\lambda^{-1}[1 - \lambda^4]^{1/2},
\label{nau}
\ee
where the donor's radius is $r_2\rsun$, and $\lambda = (M_2/M_{\rm Ch})^{1/3}$, with $M_{\rm Ch} = 1.44\msun$ the Chandrasekhar mass. This reproduces the mass--radius relation found from full structure calculations to about 2\% accuracy. The tidal lobe of the white dwarf has radius
\be 
R_L = 0.462\left(\frac{M_2}{M}\right)^{1/3}a(1 - e) 
\label{lobe}
\ee
where $M_1$ is the black hole mass and $M= M_1+M_2$ the total mass, and $a$ and $e$ are the semimajor axis and eccentricity of the white dwarf orbit. Using Kepler's 3rd law gives
\be
R_L = 4.5\times 10^{10}m_2^{1/3}P_4^{2/3}(1-e)\, {\rm cm}, 
\label{rl}
\ee
where $P_4$ is the orbital period $P$ in units of $10^4$~s. Equating this to $r_2$ in eqn (\ref{nau}) and using $m_2 = (1.44)^{1/3}\lambda$ gives 
\be
\frac{1}{\lambda^2}[1 - \lambda^4]^{1/2} = 5\times10^{10}\lambda P_4^{2/3}(1 - e)
\ee
which leads to 
\be
\lambda = \frac{1}{[1 + 527P_4^{4/3}(1 - e)^2]^{1/4}}
\ee
and so
\be
m_2 = \frac{0.013}{P_4(1-e)^{3/2}(1 + y)^{3/4}}
\label{m2}
\ee
where
\be
y = 1.9\times 10^{-3}P_4^{-4/3}(1-e)^{-2}.
\label{y}
\ee
For a circular orbit and $y \ll 1$ (which implies $M_2 \ll M_{\rm Ch}$) equation (\ref{m2}) gives the well--known mass--period relation $m_2 \propto P^{-1}$ for stellar--mass binary accretion from a degenerate companion star, as in the AM CVn systems (e.g. King, 1988). Since the range of $m_2$ is restricted, I note that long orbital periods require eccentricities very close to unity to make the pericentre distance small enough for the companion to fill its tidal lobe.

The equations for the mass transfer process driven by GW losses follow in the same way as for stellar--mass binaries (see e.g. King, 1988), but here we must allow for eccentric orbits (cf King, 2020). We compute the change of the tidal radius $R_L$ with mass transfer, and compare it with the change of the radius of the WD donor.

The orbital angular momentum of the WD--MBH system is
\be
J = M_1M_2\left(\frac{Ga}{M}\right)^{1/2}(1 - e^2)^{1/2}
\ee
where $M = M_1 + M_2$ is the total mass.
Logarithmic differentiation gives
\be
\frac{\dot J}{J} = \frac{\dot M_1}{M_1} + \frac{\dot M_2}{M_2} -{\frac{\dot M}{2M}}
+\frac{\dot a}{2a} - \frac{e\dot e }{1-e^2}.
\ee
Assuming that all the mass lost by the WD is ultimately accreted by the black hole, so that $\dot M_1 = -\dot M_2$ and $\dot M = 0$, we have
\be
\frac{\dot a}{a} = - \frac{2\dot M_2}{M_2}\left(1 - \frac{M_2}{M_1}\right) + \frac{2\dot J}{J} + \frac{2e\dot e }{1-e^2}.
\label{a}
\ee

We set $R_2 \propto M_2^{\zeta}$, where $\zeta$ is the WD radius--mass index, 
which is $-1/3$ for low $M_2$ and becomes more negative as $M_2 \rightarrow M_{\rm Ch}$. Then
\be
\frac{\dot R_2}{R_2} = \zeta\frac{\dot M_2}{M_2},
\ee
and using eqn (\ref{lobe}) we have finally
\be
\frac{\dot R_L}{R_L} - \frac{\dot R_2}{R_2} = -\frac{2\dot M_2}{M_2}\left(\frac{5}{6} + \frac{\zeta}{2} - \frac{M_2}{M_1}\right) + \frac{2\dot J}{J} - \frac{\dot e}{1 + e}.
\label{transfer}
\ee

We use this equation first to discuss the dynamical stability of mass transfer, i.e. stability on 
timescales much shorter than that for GW losses. The discussion
is essentially identical to that familar for mass transfer stability for stellar--mass binaries.
On dynamical timescales we can neglect the GW contributions to the second the third terms on the rhs of (\ref{transfer}). 
We note that
the first term on the rhs of (\ref{transfer}) is always positive, since $\dot M_2 < 0$, $M_2/M_1 \ll 1$, and from eqn (\ref{nau}) the mass-radius index $\zeta$ never reaches the value $-5/3$ required to make the bracketed term negative. This term is stabilizing, tending to expand $R_L$ away from $R_2$. 

When the system first comes into contact (i.e. the WD first fills its tidal lobe) there is a transient, slightly destabilizing (negative) contribution to the $\dot J$ term, as gas in orbit about the black hole subtracts angular momentum from the binary orbit. 

But after a viscous timescale $t_{\rm visc}$ this gas will have formed an accretion disc. Central accretion on to the black hole occurs only because viscosity transports angular momentum outwards (e.g. Lynden--Bell \& Pringle, 1974), and the disc size is limited by the WD orbit, either through tides or physical collisions. The initial orbit of gas lost from the WD is very close to its orbital radius near pericentre, so the disc reaches an effectively steady state in a few orbital periods.

At this point this negative contribution to $\dot J$ effectively stops: after this, the disc passes almost all its angular momentum back to the WD orbit, 
subtracting only the small amount given by gas accreting (with very low angular momentum) on to the the black hole at its inner edge.
This reabsorption of the original angular momentum of the mass transferred to the black hole
holds even for systems with periods 
\be
P \gtrsim P_{\rm crit} = t_{\rm visc}, \label{crit}
\ee
where it is possible that the accretion disc must re--form after only a few orbital periods. In all cases central accretion on to the black hole removes very little angular angular momentum, and the remainder must be returned to the WD orbit via tides.


Since we effectively now have $\dot J = \dot e = 0$ on short timescales in all cases,
eqn (\ref{transfer}) shows that mass transfer from the white dwarf to the black hole is always dynamically stable, contrary to the assertions in Zalamea et al. (2010) and Metzger et al.
(2022). 

The reason for stability is that the mass lost by the white dwarf and gained by the black hole is transferred towards the centre of mass of the BH--WD binary system, and therefore has lower angular momentum than before. But since we have $\Delta J=0$ on a dynamical timescale,
the binary separation has to expand (the first term on the rhs of eqn \ref{a}) to compensate for this. The result is a wider binary of longer orbital period, and slightly increased eccentricity, but with
the same angular momentum as before, because less mass is moving in this wider orbit (the distinction between angular momentum and {\it specific} angular momentum is crucial here). 
The two papers cited above claiming instability did not consider the orbital evolution forced by the mass transfer -- i.e. the first term on the rhs of eqn \ref{a}).

The conclusion here that mass transfer is stable on dynamical timescales also holds (by the same arguments) for stellar--mass binaries. The AM CVn
double--WD binary systems are examples of mass transfer from a degenerate donor star, and are well known to have highly stable mass transfer rates as their mass ratios
$M_2/M_1$ are below the value $5/6 + \zeta/2 \simeq 2/3$ required for dynamical stability (see e.g. King, 1988 for a discussion). 

This discussion shows that the effect of gravitational radiation losses $\dot J/J$ is to drive mass transfer at a rate given by setting the lhs of eqn (\ref{transfer}) to zero, specifying the mass transfer rate as
\be
-\frac{\dot M_2}{M_2}\left(\frac{5}{6} + \frac{\zeta}{2}\right) = 
-\frac{\dot J}{J} + \frac{\dot e}{2(1+e)}.
\label{mdot}
\ee
where I have used that fact that $M_2/M_1 \ll1$.
Further, since $\dot R_2/R_2 = \dot R_L/R_L$, 
the reference radius $R_2$ is always a constant multiple $\mu$ of $R_L$. The mass transfer rate is exponentially sensitive to $\mu$ because of the density stratification in the outer layers of the donor (in practice, the non--degenerate outer layers of the white dwarf atmosphere). The stable nature of the mass transfer means that the system quickly finds the required value of $\mu$, and returns to it if perturbed. 

It is important to note that the mass transfer rate computed in this way is the long--term average rate, and not the instantaneous accretion rate on to the black hole. The latter is in any case clearly observed to be variable both within eruptions and from one eruption to the next. Accretion must occur through a disc, which can vary either intrinsically, for example because of instabilities, or in response to the periodic injections of mass from the WD, which can destabilize it in various ways. But since the total mass of gas stored in the disc is limited, it is clear that the mean accretion rate deduced from an extended sequence of eruptions must match the mass transfer rate.

In King (2020) I showed that for large eccentricities $e \sim 1$ the system evolves so that $a(1-e)$ (or more accurately, $a(1-e^2)$) is almost constant, as both $a$ and $e$ decrease together. This is physically reasonable, since all the GR effects are very closely confined to pericentre, where the orbital velocity is highest, making this almost a point interaction. The mass transfer rate is 
\be
-\dot M_2\simeq
9.1\times 10^{-7}M_5^{2/3}P_4^{-8/3}
\frac{m_2^2}{(1-e)^{5/2}}
\,\msun\,{\rm yr}^{-1}
\label{grmt}
\ee
(cf eqn (15) of King 2020), 
where I have corrected the exponent of $(1 - e)$ from 7/2 to 5/2 (cf Chen et al., 2021) 
and used Chen et al.'s parametrizations $M_5 =M_1/10^5\msun, P_4 = P/(4\,{\rm hr})$. 
For $(1 - e) < 0.1$ or still smaller (see Table 1), this has the right order of magnitude 
\be
\gtrsim 3\times 10^{-4}\msun\, {\rm yr}^{-1}
\label{mag}
\ee
needed
to explain the luminosity of QPE sources (averaged on timescales much longer than the eruptions
themselves)\footnote{A full derivation from eqn (\ref{mdot}), retaining all the terms in powers of the eccentricity $e$, multiplies the transfer rate  (\ref{grmt}) by the function $f(e) = (192 - 112e + 168e^2 + 47e^3)/192$, which varies between 1 and 1.54.}.

Chen et al. (2021) show that the two constraints (\ref{m2}, \ref{grmt}) lead to the convenient forms
\be
m_2 = 0.2C^{-15/22}
\label{secmass}
\ee
\be
1-e = 0.07C^{5/11}P_4^{-2/3}
\label{ecc}
\ee
where
\be 
C = M_5^{4/15}(L\Delta t)_{45}^{-2/5}{\eta}_{0.1}^{2/5}
\label{C}
\ee
for $M_2 \ll M_{\rm Ch}$. Here $(L\Delta t)_{45}$ is the mean energy emitted
in the source's eruptions in units 
of $10^{45}\, {\rm erg}$, and $\eta_{0.1}$ the efficiency of accretion in units of $0.1c^2$.
Chen et al. (2021) show that these equations give sensible values for 
$M_2$ and $e$ for the five recognised QPE sources (see Table 1). Note that from equations (\ref{secmass}, \ref{ecc}, \ref{C}) we have 
\be
m_2 \propto M_5^{-2/11}, 1 - e \propto M_5^{4/33}
\label{Mm}
\ee
so that for otherwise fixed parameters the WD mass and eccentricity are slightly decreased for larger assumed black hole masses.

\section{The Observed Sample}

The equations derived in Section (\ref{sec:transfer}) now allow us to attempt fits to the entire observed sample of QPE sources. 
Table 1 shows the results of fitting the currently--known sample
of QPE sources. 

We can formally extend eqns (\ref{secmass}, \ref{ecc})  to all WD masses up to $M_{\rm Ch}$ by multiplying $C$ by the factor $(1+y)^{-3/5}$ (cf eqns \ref{m2}, \ref{y}). But it is straightforward to argue instead from (\ref{secmass}) that donor masses $M_2 \simeq M_{\rm Ch}$ require $C^{-15/22} \simeq 7$, and so from (\ref{C}) that 
\be 
L\Delta t \simeq 10^{48}M_5^{2/3}\, {\rm erg},
\label{Ldelta}
\ee
and from (\ref{ecc}) that
\be 
1- e \simeq 0.019P_4^{-2/3}
\label{hlecc}
\ee 
These relation are easy to understand physically: for a WD with high mass -- and so very small radius -- to fill its tidal lobe requires a small pericentre separation $\propto (1-e)P^{2/3}$, which directly gives (\ref{hlecc}). The orbital speed must be high here, so the GR evolution must be very rapid, making the mass transfer here large (cf eqn \ref{Ldelta}).
This requirement is strongly significant observationally -- the quantity $L\Delta t$ must be at least 40 times larger than any of the first five recognised QPE sources, so any system with $M_2$ approaching $M_{\rm CH}$ must have luminous and/or prolonged outbursts. 
The only reasonable candidate for a system like this 
is the unusual object 
ESO 243-39 HLX--1, often abbreviated to HLX--1. 
This system is usually regarded as the best candidate for an intermediate--mass black hole among the ultraluminous X--ray sources (`ULXs', hence the designation HLX = `hyperluminous'). Its discovery  (Farrell et al., 2009) pre--dates those of the QPEs. It has several promising features: it may be associated with a low--mass galaxy very close to a much larger galaxy (ESO 243-39), as 
studied here\footnote{This association is more plausible than postulating the presence of such massive black holes in non--nuclear regions of otherwise normal galaxies. Most ULXs are now known to be neutron stars or stellar--mass black holes fed at very high mass transfer rates. See King, Lasota \& Middleton (2022) for a comprehensive review.}.
It has had a series of X--ray outbursts repeating at intervals $\sim 1$~yr, with luminosity $L \simeq 10^{42}\,{\rm erg\, s^{-1}}$, each decaying steeply over a timescale $\Delta t \sim 10^7$s (see Lin et al., 2019, Fig 1). The first was detected in late 2008, and after 6 nearly--annual outbursts the expected 7th outburst was essentially absent. The next outburst occurred `on time', one year after the missing one, but the source then missed what would have been the 9th outburst, if all had appeared on time.

There is no clear value for the mass of the black hole, but the picture presented here requires $M_1 \geq 5\times 10^4 \msun$ if the companion mass $M_2$ is to be $< M_{\rm Ch}$. Table 1 considers this critical BH mass and gives the corresponding limit on the quantity $1 - e$. For a higher assumed BH mass 
(e.g. $10^6\msun$) the white dwarf mass is lower, but
still significantly larger than those of the shorter--period QPE sources of Table 1 unless the black hole mass is implausibly large, i.e  $M_1 \sim 3\times 10^9\msun$. 


Since HLX--1 appears to satisfy the constraint (\ref{Ldelta}) we compute $1-e$
from (\ref{hlecc}) with $P_4 \sim 2000$. This gives 
\be 
1 - e \simeq 1.2\times 10^{-4}.
\ee
(For a $10^6\msun$ black hole this extreme eccentricity is slightly reduced -- see Table 1.)
We note that as expected the very high eccentricity comes entirely from the long orbital period -- from (\ref{hlecc}) a more usual orbital period $P_4 \sim 1$ would give a fairly standard QPE eccentricity with $1 - e \sim 2\times 10^{-2}$. 

Even without the likely irregular eruption patterns, there are obvious selection effects against finding QPE systems with periods longer than HLX-1, even though they may well exist (see Section \ref{sec:discussion}). But 
it is already clear that the donor star in systems with far longer periods would remain bound to the central black hole of a low--mass galaxy. For high eccentricity the apocentre of a system like this is at a distance 
$2a$ from the black hole. Comparing this with the radius of influence $2GM/\sigma^2$ of the black hole shows that orbiting donors remain bound to the central black hole provided that their periods are less than 
\be 
P_{\rm max} \simeq \frac{8GM}{\sigma^3} \simeq 2\times 10^4M_5\sigma_{50}^{-3}\, {\rm yr}
\ee 
where
$\sigma = 
50\sigma_{50}\, {\rm km\, s^{-1}}$ is the velocity 
dispersion of the galaxy.

\begin{table} 
\centering
\caption{QPE Source Properties.  
The derived values of the donor
star mass and the orbital eccentricity for the five recognised QPE
sources plus HLX--1. 
There is no clearly established black hole mass $M_1= 10^5M_5\msun$ for HLX--1 so the results are shown for two assumed values in [\,\,]: any black hole mass 
$M_1 < 5\times 10^4\msun$ requires donor masses greater than the Chandrasekhar limit, and the values for an assumed value $M_5 = 10\msun$ are also shown for comparison (see eqn (\ref{Mm}).
(Table adapted from Chen et al., 2021.
Data from Miniutti et al., 2019; Giustini et al. 2020; Song et al. 2020;
Payne et al. 2020;
Arcodia et al. 2021; Chakraborty et al. 2021.
}
\hspace*{-1cm}\begin{tabular}{l c c c c c} 
source & $M_5$ & $P_4$ & $(L\Delta t)_{45}$ & $m_2$ & $1-e$ \\
GSN 069 & 4.0 & 3.16 & 10 & 0.32 & $2.8\times 10^{-2}$  \\
RX J1301.9 
& 18 & 1.65 & 1.7 & 0.15 & $7.2 \times 10^{-2}$\\
eRO -- QPE1 & 9.1 & 6.66 & 0.045 & 0.46 & $1.4\times 10^{-2}$\\
eRO -- QPE2 & 2.3 & 0.86 & 0.80 & 0.18 & $9.9\times 10^{-2}$ \\
XMMSL1 
& 0.85
& 0.90 & 0.34 & 0.18 & $9.9\times 10^{-2}$\\
HLX--1 
& $[0.5]$
& 2000 & 1000 & 
$ 1.43$
& $ 1.2\times 10^{-4}$\\
 \,  . . .  
& $[10.0]$ & . . .   & . . .  & 0.81 & $1.5\times 10^{-4}$ \\

\end{tabular}

\end{table}

\section{X--ray Light Curves} \label{sec:lc}

The X--ray emission of the QPE sources must be powered by accretion on to their black holes. This occurs through an accretion disc, which in general must be warped, as the plane of the WD orbit is unlikely to
coincide with the spin plane of the black--hole accretor.
The quasiperiodic nature of the X--ray light curves evidently reflects the reaction of the accretion disc to the periodic interaction with the WD. This is specified by the viscous timescale of the accretion disc
\be
t_{\rm visc} \simeq \frac{1}{\alpha}\left(\frac{R}{H}\right)^2\left(\frac{
R_d^3}{GM_1}\right)^{1/2}
\ee
where $\alpha \sim 0.1$ is the Shakura--Sunyaev viscosity parameter, and $R_d$ is the disc radius, which we estimate from the tidal condition (\ref{lobe}) as
\be
R_d = 2.5\left(\frac{M_1}{M_2}\right)^{1/3}R_2.
\ee
This gives
\be
t_{\rm visc} \simeq \frac{4}{\alpha}\left(\frac{R}{H}\right)^{2}\left(\frac{R_2^3}{GM_2}\right)^{1/2}.
\ee
The dynamical time $(R_2^3/GM_2)^{1/2}$ of a WD is $\sim 1$~s, so we conclude that 
\be
t_{\rm visc} \sim \frac{4}{\alpha}\left(\frac{R}{H}\right)^2\, {\rm s}.
\label{tvisc}
\ee
We can apply this result to two aspects of QPE sources. 

First,
we noted above (eqn \ref{crit}) that for orbital periods 
$P \gtrsim t_{\rm visc}$
the accretion disc may have to re--form after a few orbital periods, so we might expect an irregular light curve, with `missing' eruptions, at such periods. For these wide systems, the disc is undisturbed by the orbiting WD except for very brief interludes, so its aspect ratio $H/R$ should be close to the value
$H/R \sim 10^{-3}$ (e.g. Collin--Souffrin \& Dumont, 1990) expected for an extended disc around a supermassive black hole. This gives
\be 
P_{\rm crit} \sim 1\, {\rm yr}.
\label{pcrit}
\ee

Encouragingly, this agrees with the irregular eruption behaviour of HLX--1 ($P \sim 1\, {\rm yr}$) noted above.
This source has defied a number of attempts to model its light curve in terms of of accretion disc instabilities (e.g. Lasota et al., 2011), and various other suggestions as to its unusual nature (e.g. King \& Lasota, 2014). The discussion here suggests instead that the accretion disc runs out of gas after a few orbits and has to re--form.



Giustini et al (2021) noted that the QPEs from RX 
J1301.9+2747 showed an alternating pattern of long and 
short recurrence times, and that GSN~069 also 
shows this behaviour in a milder form. Since then it 
has become clear that some of the five recognised QPE
sources show this pattern from time to time, but the
correlations between waiting times and amplitudes are 
complex (see Fig. 3 of 
Chakraborty et al. (2021)). 

A distinctive feature of QPE sources is that their mean accretion rates $\dot M = L\Delta t/P\eta c^2$ imply remarkably high ($\sim 10^{-5}\msun\,{\rm yr}^{-1}$) mass transfer rates from 
the white dwarf donors compared with those in stellar--mass binaries. They imply changes $\Delta R_L$ in the Roche lobe radius of order $\sim 10^{-5}R_L$ per year. This is comparable to the atmospheric scaleheight $H = kT/\mu m_H g$ (where $T$ is the WD surface temperature and
$g = GM/R_2^2 \sim 10^8\, {\rm cm\, s^{-2}}$ the surface gravity), which gives 
$H \sim (10^{-5} - 10 ^{-4}) R_L$ for surface temperatures $T \sim 10^4 - 10^5$~K. 

Then if the star remains close to hydrostatic balance
the instantaneous mass transfer rate must average to the long--term evolutionary mean (given by the GR angular momentum loss) on timescales $\lesssim 1$~yr. This is already extremely short compared with most mass--transferring binaries of stellar mass, where the evolutionary average is only enforced on unobservably long timescales (this is the reason why observed period derivatives for most mass--transferring systems such as CVs do not in general agree with the expected long--term evolutionary rate, but show a very large scatter instead). 

But the mass transfer timescale in the QPE sources is likely to be even shorter than the hydrostatic value derived above. Mass transfer only occurs near pericentre, so the Roche lobe closes in on the star and then out again dynamically on each orbit. The star is then essentially an oscillator being forced at a frequency close to resonance, so its radius response must be significantly larger and faster than the quasistatic one considered above. 

A
full fluid--dynamical treatment is required to calculate this
near--resonant forcing, but it is already clear that this can have major effects on QPE light curves. In some cases it must force the mass transfer rate to average to the evolutionary mean over a very few binary periods. This probably accounts for the long--short behaviour seen in several QPE systems: a burst of mass transfer well above the evolutionary mean expands the tidal lobe so far above the stellar surface that the next burst must be far weaker. This undershoot then leads to a longer burst next time, and so on. 

Such alternating episodes probably change the shock conditions 
where the gas from the WD interacts with the disc, and so its aspect ratio $H/R$. The estimate (\ref{tvisc}) then shows that the effects of the differing bursts of mass transfer must affect the timescales for delivering gas to the vicinity of the black hole where the X--ray emission is produced. This may be why the eruption times deviate from strict periodicity, but evidently a full hydrodynamic calculation is needed to answer this question.




This suggests that the distinctive properties of several short--period QPE light curves could follow from the fact that their mass transfer rates are high, but stable. The long orbital period of HLX--1 may allow any near--resonant oscillations to damp more between pericentre passages, but significant changes in its light curve may also result from the extreme Einstein precession produced by the very high eccentricity. Periastron passage here is effectively scattering through a large angle.

\section{Orbital Evolution}

The orbits of QPE binaries evolve in time because gravitational radiation extracts both angular momentum and energy in significant amounts. For zero eccentricity the corresponding pair of equations describing the evolution give the same information, but for QPE sources the significant eccentricities mean that we need both equations. From the WD mass--period relation (\ref{m2}) we see that
\be
\frac{\dot P}{P} = -\frac{\dot M_2}{M_2} + \frac{3\dot e}{2(1-e)},
\label{evol}
\ee
where the first term (the effect of mass transfer expanding the orbit) is positive, while the second (circularization) is negative. (As noted above, for a circular orbit this term vanishes, so the period increases.)


To find an expression for the period derivative we cannot simply use the equation for $\dot e$ from Peters (1964) as this does not allow for mass exchange. Instead we use the fact that for constant total binary mass $M = M_1 + M_2$ the period is always $\propto a^{3/2}$ by Kepler's law. To get $\dot a$ we note that the energy of the binary orbit is
\be
E = -\frac{GM_1M_2}{2a}
\ee
so that for $M_1 \gg M_2$ we have
\be
\frac{\dot a}{a} = - \frac{\dot E}{E} + \frac{\dot M_2}{M_2}
\label{aE}
\ee{}
We use (\ref{mdot}) to eliminate $\dot M_2$ in favour of $\dot J$, 
and Peters' equation for the GR energy loss,
giving eventually
\be
\frac{2\dot P}{3P} = \frac{\dot a}{a} = -\frac{2f(e)}{\tau},
\label{pdot}
\ee
where 
\be
\tau = \frac{5c^5a^4(1-e^2)^{7/2}}{32G^3M_1M_2M}
\label{time}
\ee
is the gravitational wave timescale, and for $\zeta \simeq -1/3$
\be 
f(e) = \frac{3}{2} + \frac{143}{48}e^2 - \frac{5}{96}e^4.
\label{quad}
\ee{}
It is now straightforward (although tedious) to verify that all of $a, P$ and $e$ decrease on the timescale $\tau$, since we can combine (\ref{quad}) with eqn (\ref{a}) to get the evolution of the eccentricity $e$ when mass is exchanged -- (cf eqn \ref{ecc}).

The orbital evolution of QPE sources towards more circular binaries with shorter orbital periods is qualitatively similar to that of extreme mass ratio inspiral events (EMRIs). The mass transfer rates increase over time, but the systems may become less recognisable as
QPE sources because the decrease in $e$ means that mass transfer is spread more evenly over the orbital period. Despite this, it is evidently worthwhile studying X--ray emission from galaxy nuclei at high time resolution with the aim of finding such systems.

\section{Discussion} \label{sec:discussion}

There are several suggested models for QPE sources. Ingram et al. (2021)
note that a double black--hole binary (forming as the result of a merger) observed edge--on, may produce the observe flares through gravitational lensing of an accretion disc around one of the holes. 
This model can in principle explain sharp and symmetrical light curves, but observed QPEs are often more messy than this. In addition the model has difficulty simultaneously explaining both the amplitude and duration of the flares. Further, gravitational lensing is achromatic, whereas QPEs look different at different energies -- for example shorter at hard X--rays than soft.

Xian et al. (2021) suggest that QPEs arise from collisions of an orbiting star with a central black hole accretion disc. This is potentially attractive in offering a possible explanation for the alternating behaviour seen in the first observed QPE light curves, but as we noted above, much more complex patterns appear in QPE sources found later. In addition, there is no obvious reason why this mechanism cannot apply in galaxies with more massive black holes, unless the star--disc collisions are somehow systematically too faint compared with the central accretion luminosity, so the bias towards low--mass black holes is unexplained. Moreover, not all QPE sources are in otherwise active galaxies, so it is not obvious that the central black hole has a well--developed accretion disc in every case.

The observed bias towards low black hole masses  suggests a connection with EMRI events. Metzger et al. (2022) suggest a picture where two simultaneous EMRIs 
share the same orbital plane, and produce the QPEs through mutual gravitational interactions. This is inherently a less likely event than a single strongly eccentric orbiter, but is adopted because of a belief that mass transfer from a single white dwarf filling its tidal lobe in such an orbit is unstable. Section 2 above shows that this assertion is incorrect.

The work of this paper strengthens the case that QPEs are a result of periodic mass transfer from orbiting low--mass stars which narrowly escaped full tidal disruption. Observational selection means that we currently can only see those cases where the donor star is a white dwarf. This agrees with the CNO--processing seen in the spectrum of GSN~069,
and it gives detailed fits to the data on the first five recognised QPE sources, as well as the much longer--period system  HLX--1. The predicted lifetimes are short in all cases.  The accretion luminosity and WD mass for GSN 069 give about 3200 yr. The longest lifetime is 
$\sim 2\times 10^5$~yr for HLX--1.

These short lifetimes suggest that the events producing these systems must be fairly frequent, and so that they make a significant contribution to the growth of the central massive black hole. It seems inevitable that there must be many more events involving main--sequence donors which we cannot directly identify. 
For these systems the mass--radius relation $R_2 \propto M_2$ gives the mass--period relation as 
\be
m_2 = 0.26P_4(1-e)^{3/2}
\ee
instead of (\ref{m2}). Using this to eliminate the orbital period from (\ref{grmt}) 
gives
\be 
-\dot M_2({\rm MS}) = 2.6\times 10^{-8}M_5^{2/3}m_2^{-2/3}(1 - e)^{3/2}\, \msun\,{\rm yr}^{-1}.
\label{MS}
\ee 
Using (\ref{m2}) in (\ref{mag}) gives
\be
-\dot M_2({\rm WD}) = 2.1\times 10^{-4}M_5^{2/3}m_2^{14/3}(1 - e)^{3/2}\, \msun\,{\rm yr}^{-1}.
\ee

Thus MS stars contribute much lower emission than white dwarfs of the same mass $M_2$ for a given black hole mass and eccentricity, unless $M_2 <0.19\msun$.

The formation mechanism for these systems is not yet clear. King (2020) suggested that the low mass of the WD in GSN 069 was more easily understood as a result of disrupting a low--mass giant than direct capture, but it appears problematic to achieve the observed tight orbits for QPE sources in this way, and we have in any case seen that more massive WDs are present  in QPE sources with longer orbital periods.

Recently Cufari et al. (2022) pointed out that capturing a main--sequence star into a QPE binary in a single scattering event is difficult, as this would dissipate more than the star's binding energy. They suggest instead that
formation is possible by the Hills (1988) mechanism, where a close stellar--mass binary is `ionized' by the MBH, one member being gravitationally captured by the MBH, and the other ejected as a hypervelocity star. 
On the other hand introducing a WD companion directly by single scattering is allowable, as the binding energy of a WD is far higher than for a solar--type star. In line with this, the evolution of QPE binaries discussed here -- particularly the parameters derived in Table 1 -- suggests that the observed QPE systems descend from similar but relatively unobservable systems with more massive WDs in more eccentric long--period orbits.



\section*{Data Availability}

No new data were generated or analysed in support of this research.

\section*{Acknowledgments}
I thank the authors of Cufari et al. (2022) for sending me a copy of their paper in advance of publication. I am particularly  grateful to the anonymous referee for a very thoughtful and perceptive report.




\end{document}